\documentclass[runningheads]{llncs}
\usepackage[T1]{fontenc}
\usepackage{graphicx}
\usepackage[hidelinks]{hyperref}
\usepackage{color}

\usepackage{subcaption}
\begin{document}
\title{Not All Trust is the Same: \\ 
Effects of Decision Workflow and Explanations in Human-AI Decision Making}
\titlerunning{Not All Trust is the Same}
%

\author{Laura Spillner\orcidID{0000-0001-8490-8961} \and
Rachel Ringe\orcidID{0009-0005-4696-5873} \and \\
Robert Porzel\orcidID{0000-0002-7686-2921} \and
Rainer Malaka\orcidID{0000-0001-6463-4828}}
\authorrunning{L. Spillner et al.}
%
\institute{Digital Media Lab, University of Bremen, Germany\\
\email{\{l.spillner,rringe,porzel,malaka\}@uni-bremen.de}}

\maketitle              
\begin{abstract}
A central challenge in AI-assisted decision making is achieving warranted, well-calibrated trust. Both overtrust (accepting incorrect AI recommendations) and undertrust (rejecting correct advice) should be prevented. Prior studies differ in the design of the decision workflow - whether users see the AI suggestion immediately (1-step setup) or have to submit a first decision beforehand (2-step setup) -, and in how trust is measured - through self-reports or as behavioral trust, that is, reliance. We examined the effects and interactions of (a) the type of decision workflow, (b) the presence of explanations, and (c) users’ domain knowledge and prior AI experience. We compared reported trust, reliance (agreement rate and switch rate), and overreliance. Results showed no evidence that a 2-step setup reduces overreliance. The decision workflow also did not directly affect self-reported trust, but there was a crossover interaction effect with domain knowledge and explanations, suggesting that the effects of explanations alone may not generalize across workflow setups. Finally, our findings confirm that reported trust and reliance behavior are distinct constructs that should be evaluated separately in AI-assisted decision making.

\end{abstract}

\section{Introduction}



AI systems increasingly act as everyday assistants, for drafting texts, for taking on tasks as agents, and even to advise on decisions that affect people’s lives. Yet, these systems often confidently assert incorrect information \cite{xiong_can_2023} or hallucinate output \cite{huang_survey_2025}. Users are thus faced with the question of when they should trust AI advice. In \emph{AI-assisted decision making}, AI decision support systems (DSS) provide a recommendation for a decision based on the AI's prediction, while the human user has the final say. Trust, then, is a double-edged sword - users might overrely on incorrect AI recommendations or reject accurate ones. The challenge is not to maximize trust but rather to calibrate it: users should trust the AI advice when it is correct and resist it when it is wrong. 
\emph{Calibrated trust} is achieved when ``the perceived trustworthiness of a system matches the actual trustworthiness of a system'' \cite[p. 2]{wischnewski_measuring_2023}. In this context, trust can be understood either as users' perceived, reported trust, or as trusting behavior, i.e., following or rejecting AI advice, which is often referred to as \emph{reliance} \cite{lai_towards_2023}. 

Previous research has explored different ways to support trust calibration by making system reliability and individual predictions more transparent, e.g., through explanations, which can aid the user in their decision to reject or follow the AI suggestion \cite{lai_towards_2023}. While explanations seem promising to help calibrate user trust, results on their effectiveness to do so have been mixed \cite{wang_are_2021}.
Another proposed approach is to alter the decision workflow - for instance, requiring users to make an initial choice before seeing the AI’s recommendation (a 2-step setup) rather than being shown it immediately (a 1-step setup). This \emph{cognitive forcing} design might reduce \emph{overreliance}, that is, the rate at which users accept incorrect AI recommendations \cite{bucinca_2021}. However, it remains unclear if such an effect holds for reported trust, behavioral trust, or both; and how the effect of the decision setup interacts with the presence of explanations \cite{bucinca_2021}.


To address this gap, we investigated how decision workflow and explanations interact to influence users’ trust in and reliance on an AI DSS. 
We conducted a between-subjects online study ($N = 300$) presenting users with a fictional workplace scenario at a university, in which they had to make a decision with AI assistance, which would affect the academic careers of university students. 

The results of this study showed, firstly, that self-reported trust and behavioral trust were weakly correlated; thus, the two are related, but should be treated as two distinct constructs. Secondly, we could not confirm the potential of a 2-step setup to improve trust calibration - in the given task domain, it increased overreliance. Moreover, while it did not directly impact reported trust, we observed a crossover interaction effect with the presence of explanations: explanations were only helpful in increasing trust in the 2-step setup. This shows that results concerning the impact of explanations on user trust cannot be generalized between different decision workflows. 
Our study contributes empirical evidence on how design choices in AI-assisted workflows shape trust and overreliance, based on which we suggest concrete implications for DSS design. 

\section{Related Work}

Li et al. propose that AI DSS can be divided into three broad categories: AI as advisor (which is what ``AI-assisted decision making'' usually refers to), human as supervisor, and collaborative setups \cite{li_as_2025}. 
In AI-assisted decision making, the human decision maker integrates the AI prediction (and potentially its explanation) with their own understanding of the situation to make a final decision. This can enhance human capabilities, but it is not suitable when human performance is much weaker than AI accuracy. It poses the challenge of designing a system that users will trust, without creating overtrust that leads them to disregard their own intuition. In contrast, if the human acts as a supervisor, they provide oversight for a decision made by an AI. This is more prevalent in domains where AI has better accuracy than humans, but humans can observe and intervene in case of anomalies. Thus, the main challenge is how to help users recognize these, without creating undertrust that would lead to AI accuracy being underutilized. In a collaborative process, human and AI act as peers: this setup is less susceptible to over- or undertrust, but requires being able to compare their respective decision suggestions equally, e.g., by making a final decision based on the respective task confidence of the user and the AI. Our study focuses on the area of AI-assisted decision making.

\emph{Effective human-AI collaboration} is achieved when the performance of the human-AI team surpasses that which either human or AI can achieve on their own. In some cases where team performance is better than that of humans alone, this is only because the AI system itself has a much higher accuracy than humans \cite{zhang_effect_2020}. Ideally, the user's and the system's respective knowledge complement each other, meaning that humans have some intuition that is different from that of the AI \cite{chen_understanding_2023}. This has been termed \emph{complementary collaboration} \cite{bansal_does_2021,ma_who_2023}. 

To achieve greater team than individual performance, users' trust in the AI should be calibrated \cite{jacovi_formalizing_2021,zhang_effect_2020,wischnewski_measuring_2023}. 
Wischnewski et al. reviewed how trust calibration has been evaluated in prior studies \cite{wischnewski_measuring_2023}: On the one hand, trust calibration can be understood on a global system level to mean that users' trust in a more trustworthy (e.g., more reliable or accurate) system should be higher than their trust in a less trustworthy one. For a single system with unchanged trustworthiness, however, evaluating trust calibration is difficult. A common approach is to evaluate differences in behavioral trust, e.g., in terms of appropriate reliance (the rate of following correct advice and rejecting wrong advice), overreliance (accepting incorrect advice), or underreliance (rejecting correct advice) \cite{jacovi_formalizing_2021,wang_effects_2022}.


\subsection{Measuring Trust}


When users have to decide whether or not to trust the AI on a single decision (e.g., one task in a series of decision tasks), they can utilize information provided by the system, such as its confidence \cite{zhang_effect_2020,rechkemmer_when_2022,li_as_2025} or an explanation of the prediction \cite{chen_understanding_2023}. Some prior studies have found that this improves trust calibration \cite{poursabzi-sangdeh_manipulating_2021}, however, results are contradictory: in some cases, confidence or explanations increased overreliance \cite{suresh_misplaced_2020,zhang_effect_2020,chen_understanding_2023}, did not impact trust and thus did not improve trust calibration \cite{kunkel_let_2019}, or did not help users in understanding the system better \cite{wang_are_2021}. This disconnect could be due to differences between domains, in the task familiarity of participants, or in their general disposition towards trusting AI \cite{poursabzi-sangdeh_manipulating_2021}.
User characteristics appear to be a primary driver of trust in AI systems \cite{bach_systematic_2024}. One factor is familiarity with AI: Wazzan et al. showed that users' AI literacy impacts how they engage with explanations and how much they follow AI recommendations \cite{wazzan_evaluating_2025}.
Therefore, user characteristics and domain factors should be considered when studying AI-assisted decision making \cite{lai_towards_2023}.

Measuring trust in an AI system can be achieved through behavioral measures or by relying on self-reports through questionnaires \cite{rechkemmer_when_2022,lai_towards_2023}. Overviews with examples for both kinds of measurements can be found in \cite{lai_towards_2023} and \cite{wischnewski_measuring_2023}. 
Reported trust measures vary between ad-hoc 1-item questionnaires 
and validated ones like the HCT questionnaire \cite{madsen2000measuring}, the Trust in Automation (TiA) questionnaire \cite{korber2019theoretical} or the Trust Scale \cite{hoffman2018metrics}. 
The most common behavioral measures are agreement with the AI's suggestion (agreement rate) or changing one's decision if the AI suggestion disagrees with the user's first choice (switch rate) \cite{wang2021explanations,rechkemmer_when_2022}. 

Many researchers have cautioned that behavioral measures and reported trust may be measuring different constructs \cite{lai_towards_2023,rechkemmer_when_2022}. Some authors use \emph{reliance} to refer to behavioral measures, especially the agreement rate \cite{lai_towards_2023,xiong_can_2023,he_is_2025}, while others use ``reliance'' more generally to refer to trust in automated systems \cite{wischnewski_measuring_2023}. In the following, we will use reliance in the behavioral sense. Reliance measures give an indication of how much users will follow the system's advice in practice, which might be influenced both by perceived trust as well as by other factors such as the perceived risk of the decision or users' self-confidence \cite{lai_towards_2023}. Few studies have investigated the extent to which reported trust and reliance are correlated or impact each other. 
While some have found a weak but significant correlation between the two, supporting the notion that perceived trust affects reliance \cite{schaffer_i_2019}, some have found that subjective measures could not be used as a predictor for behavioral trust \cite{zhang_effect_2020} or performance \cite{bucina_proxytasks}. Others have encountered a \emph{trust-reliance-paradox}, that is, a mismatch between users' reported trust and reliance~\cite{schmitt2021towards,spillner_eurovision}. 

There remains a gap in establishing to what extent reported trust and reliance are related, especially across different domains, systems, and task setups. Our research thus contributes to a better understanding of this relationship for different decision making setups.

\subsection{Reducing Overreliance Through the Decision Workflow}

The effectiveness of explanations to calibrate trust and reduce over- and underreliance differs greatly based on technical features and user and domain characteristics. Moreover, untrustworthy AI can abuse the same effects to deceive users into trusting it \cite{banovic_being_2023}. 
Buçinca et al. suggest that cognitive forcing mechanisms might be used to improve trust calibration by reducing overreliance \cite{bucinca_trust_2021}. These included introducing a delay before the user can see the AI suggestion, letting them decide whether or not they want to see the AI recommendation, or making the user submit their own choice before seeing the AI recommendation, and then making a final choice. This 2-step setup also makes it possible to measure the rate at which users change their mind (switch rate), which has been used as a behavioral reliance measure in previous studies \cite{he2023knowing,wang2021explanations,rechkemmer_when_2022,lai_towards_2023}. 

\begin{figure}
\centering
\includegraphics[width=0.99\textwidth]{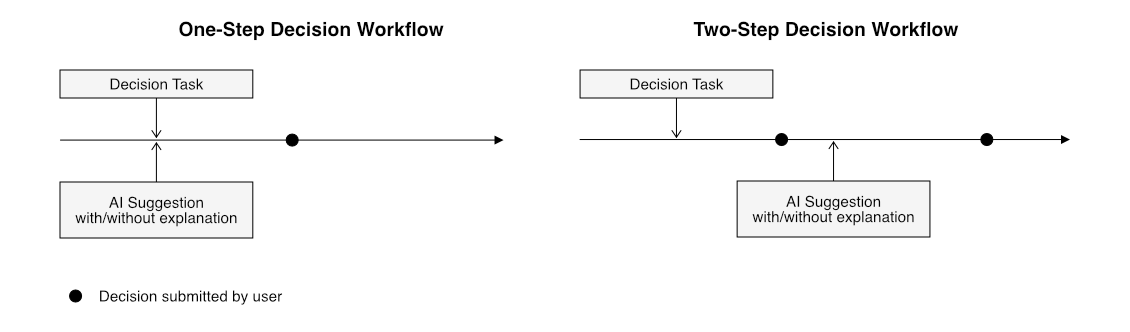}
\caption{Comparison between 1-step and 2-step setup.} 
\label{fig:workflow}
\end{figure}

While a number of prior works on human-AI decision making have used a 2-step workflow (see Figure \ref{fig:workflow}) in their studies, understanding of its effects remains limited, as only few have directly compared it to a 1-step setup. 
One such study was conducted with veterinary radiologists to explore how the number of steps in the decision workflow influences the performance of the human-AI team and the agreement rate with the AI. The results showed that, when provided with an AI recommendation and a confidence score, participants in the 2-step workflow group were less likely to agree with the AI system \cite{fogliato2022}. In another study, the 2-step setup was investigated in comparison to a 1-step workflow with an explanation, and the results showed that the 2-step workflow improved performance of the human-AI team \cite{green_2019}. Bu\c{c}inca et al. conducted a similar study that compared the 2-step setup to a standard 1-step setup with an explanation, and found that the 2-step setup did reduce overreliance more than the explanation approach, and led to generally lower reported trust \cite{bucinca_2021}. 

Although these prior studies have shown the potential of a 2-step workflow as a mechanism in AI DSS design, two of them have focused on performance as the outcome variable, and it remains unclear how the effects of the setup might interact with the effects of explanations. Our study contributes to this research by evaluating the effect of the decision workflow on both reported and behavioral trust (including overreliance), and analyzing how it interacts with the presence of explanations.

\section{Study Design}

%


The goal of this study was (a) to better understand the relationship between reported and behavioral trust measures for human-AI decision making, and (b) to investigate if adopting a 2-step decision setup impacts reported or behavioral trust in an AI DSS. Therefore, we pose two research questions:

\begin{itemize}
    \item \textbf{RQ1:} To what extent do self-reported trust in an AI system and reliance on the AI's suggestions (in terms of agreement rate) measure the same underlying construct of trust?
    \item \textbf{RQ2:} Does using a 2-step decision setup impact self-reported trust or reliance compared to a 1-step setup? 
    \begin{itemize}
        \item Does the effect of the decision setup interact with the presence of explanations?
        \item Does the effect of the decision setup interact with user characteristics such as domain knowledge or prior AI experience?
    \end{itemize}
\end{itemize}
    

\subsection{Experimental Design}

To investigate these questions, we conducted a between-subjects online study (N = 300). Participants were presented with a series of binary decision tasks involving the use of an AI DSS in a fictional workplace scenario. 

\subsubsection{Independent Variables}

The study used a two-by-two between-subjects design to compare the different interaction setups: There were two decision workflows, the 1-step and the 2-step setup, as well as one version in which the AI explained its prediction and a baseline version where the prediction was presented without explanation. In addition, we asked participants how they perceived their own knowledge of the domain (on a five-point Likert scale), and whether or not they had prior experience with using AI systems, as both of these have been previously suggested to impact user trust in AI or to interact with the effect of explanations or other features of the DSS \cite{schaffer_i_2019,suresh_misplaced_2020,wang_are_2021,wazzan_evaluating_2025}

\subsubsection{Dependent Variables}

We measured participants' trust in the AI DSS as: 

\begin{itemize}
    \item \textbf{Self-reported trust} through a modified version of the Human Computer Trust (HCT) questionnaire.
    \item \textbf{Behavioral trust}, that is, \textbf{reliance}, was measured in three ways:
    \begin{itemize}
        \item through the overall agreement rate
        \item overreliance: the rate at which participants accepted incorrect AI advice
        \item for the groups in the 2-step setup, through the switch rate: the rate at which participants changed their mind in those cases where the AI suggestion disagreed with their first choice.
    \end{itemize}
\end{itemize}





\subsection{Domain and Task}

Lai et al. provide recommendations for the choice of the decision task \cite{lai_towards_2023}. Firstly, the potential effect or risk of the decision might impact users' willingness to trust the AI system or themselves, with high-stakes tasks being more critical to assess. Secondly, if the AI performs much better at the task than most humans (or vice versa), it becomes trivial to always accept either the AI suggestion or one's own intuition - to see an improvement in collaborative decision making, knowledge has to be complementary. Thirdly, the domain knowledge of participants matters. They note that there is currently a misbalance between different domains based on the availability of suitable datasets, with little research having been conducted on professional tasks in human resources, even though this is an area where automated decisions are likely to be utilized \cite{lai_towards_2023}. 

We decided to present participants with a decision task related to the academic career outcomes of university students\footnote{The dataset is available \href{https://www.kaggle.com/datasets/thedevastator/higher-education-predictors-of-student-retention}{here via Kaggle}.} \cite{realinho2022predicting}. They were asked to imagine that they worked in a student support office, and were shown several data points about a series of students currently in their second year. These included the students' academic success so far, as well as additional demographic data about them. Based on these, they had to decide whether the students were on track to graduate successfully, or were in danger of dropping out - in the latter case, the students could be allocated additional support resources. To help make their decision, participants were provided with an AI prediction for each student's academic career outcome, but were to make the final decision by themselves.

The data for each student includes, firstly, their success in the first two semesters as grades and number of exams taken and passed so far, and secondly, demographic data including family background, history of study and degrees earned, and information about tuition payment and debt. The dataset also provides the ground truth for whether the student graduated or dropped out. 

This task was chosen because it provides a reasonably realistic context in which an AI-assisted decision would impact real human lives. As it is relatively higher-stakes without being highly critical, such as medicine, we considered that it would be reasonably similar to currently underresearched domains like human resources. It is also based on domain knowledge about the university system and grades, for which it is easy to find participants with prior experience. We performed a pre-study to test the adequacy of this decision task in terms of participants' domain knowledge, and found that humans sampled from our University workplace and personal network performed at an accuracy of 70\% - a level which makes it neither too easy to need the AI assistance, nor so hard that it is almost always rational to follow the AI advice.

\subsection{Simulated AI System}

In order to achieve a situation where user and AI can have complementary knowledge and both correct and incorrect predictions are encountered by the participants, the experimental AI system in an `AI as advisor' setup should have a similar, or only slightly higher, level of accuracy on the task than the study participants. However, while the dataset we tested was well-suited for the human participants, machine learning models can easily achieve very high accuracy ($> 95\%$) even when limiting training data, providing only a few of the available features, and using simple models. Because of this limitation, and the fact that the same task in a realistic setting might be much less clear-cut compared to the dataset, we decided to simulate the experimental AI system in a Wizard-of-Oz setup instead of using a real model. In this way, we could artificially constrain the system to an accuracy of 73\%: 15 decision cases were randomly chosen from the dataset for the study, and four out of those were answered incorrectly by the AI. The AI answers (correct or incorrect) were randomly chosen before the study, and the AI explanations were precomputed by a rule-based system (there was no human `wizard' simulating the AI). 

Such Wizard-of-Oz setups have been used in other human-AI decision making studies \cite{bucinca_2021,binns_reducing}, as is also noted in the review by Lai et al. \cite{lai_towards_2023}, however, the authors of this review caution that such a setup might not simulate errors and explanations realistically, and should only be used when the study design can still sufficiently approximate the model behaviors that are of interest. In this case, since we were not interested in the effects of different kinds of explanations or realistic model behaviors, but instead in the impact of the decision task setup itself, we decided that it was worth the compromise to use this setup to study the proposed task while keeping human and AI performances aligned. 

One group of participants was presented with the AI predictions without any further context. The other group was shown a simple local feature importance explanation, in order to assess possible interactions between the decision setup and the presence of explanations. These explanations were created by calculating how strongly each of the features predicted the outcome of a student either graduating or dropping out, comparing these with the values of the features for a given case, and choosing those three features that, in the given case, aligned the most with the prediction. In this way, we approximated what a local salient feature explanation for the simulated AI model might look like. When presenting the explanation, the three most pertinent features were listed and highlighted in green or red in the table with the student data.

\subsection{Online Study}

The study was conducted online with 300 participants using a web application developed with the Python package Streamlit
. Participants were recruited using the platform Prolific 
and were randomly assigned to one of the four groups in the 2x2 procedure. They received \textsterling 3 for their participation. 
The mean completion time was 15.07 minutes. Two participants were automatically excluded by Prolific as they experienced a timeout after exceeding the 60-minute mark calculated by Prolific - they were not compensated, and their data was excluded.

No screening was applied to the participants to obtain a random sample with regard to age and nationality, but a balanced sample was chosen with regard to gender. We expected that a random sample of Prolific users should include a reasonable percentage of participants with knowledge in the domain of University studies. Indeed, two-thirds of participants agreed or strongly agreed that they had the required domain knowledge to solve the tasks successfully, and only 29 of the other participants explicitly disagreed with that notion. 108 participants reported being students themselves.

After receiving a general introduction on the length of the study, data collection, and the scenario they were to imagine, participants were shown a series of 15 decision tasks. For each  task, they were presented with a table of student data from the academic career prediction dataset, and, because the dataset was for Portuguese students, information on how to convert grades to US grades. The participants were not told whether or not their decision had been correct after each case, in order not to influence their trust in the AI in that way. All participants were shown the same student cases, but in randomized order. 

\begin{figure}
\centering
\includegraphics[width=0.8\textwidth]{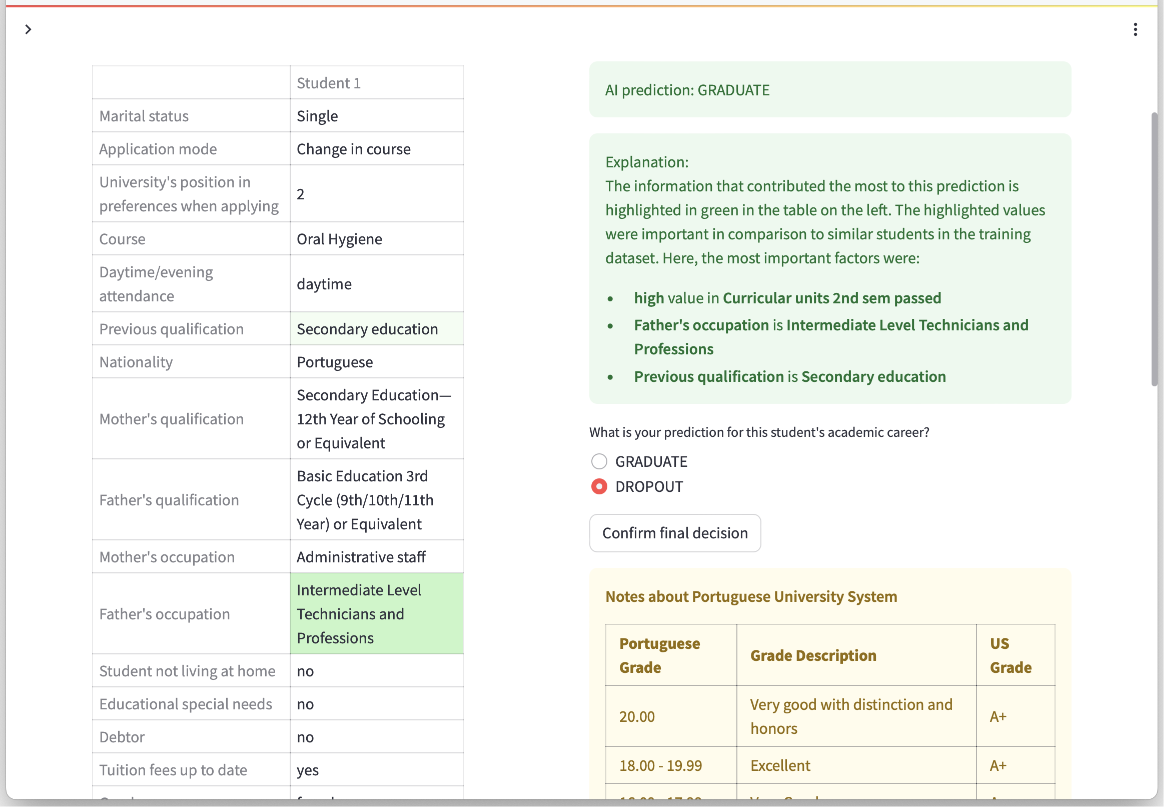}
\caption{The DSS interface. In the groups without explanations, only the prediction itself (`graduate' or `dropout') was shown, without the explanation box underneath or the colored highlighting in the data table on the left.} 
\label{interface}
\end{figure}


After making a decision for all 15 student cases, participants were asked to fill out a questionnaire. This included a modified version of the HCT questionnaire \cite{madsen2000measuring}. Compared to the original, we omitted items 3 and 4 from the `Perceived Reliability' section (those related to different conditions of use, which were not something the participants experienced); and item 4 from the `Perceived Technical Competence' section (this relates to information entered by the user, which was also not the case). We also did not use the additional sections on understandability (which focuses on whether the user understands how to operate the system - this would be relevant for a more complicated interface) and personal attachment (which relates to users' attachment to the system - this is over more extended periods of use). This modified version of the HCT had been used in a previous study on human-AI decision making \cite{spillner_eurovision}.
To assess possible covariates such as domain knowledge and AI experience, a short additional questionnaire asked participants if they agreed that they had the required domain knowledge to fulfill the task (on a 5-point Likert scale), whether they had previous experience with AI, and if so, in what context. At the end of the study, participants were informed how many of their choices had been correct.

\section{Results}

Out of 298 participants who completed the study, 152 used the 1-step decision setup (74 with explanations and 78 without), and 146 used the 2-step decision setup (74 with explanations and 72 without).
They were between 19 and 71 years old ($M = 29, SD = 9.2$ years). 
Participants originated from 29 different countries around the world, with most living in Europe (152), Africa (105), or the Americas (29), and 11 participants from Asia and Oceania (one unknown). 

Overall, performance on the decision task was in alignment with the pre-study: Final accuracy was 73\%, and, for those participants in the 2-step decision setup, their accuracy before seeing the AI suggestion was 67\%. As final accuracy equals AI accuracy, in this case, there was not enough complementary knowledge to lead to better team performance than individual AI performance. 

Participants took on average 15.07 minutes to complete the study, with 7 seconds per task. We noticed that participants in the 2-step decision setup took more time on average (M=16.4 minutes, SD=10.8 minutes) than those in the 1-step decision setup (M=13.8 minutes, SD=8.0 minutes) to complete the study. 

\subsection{Trust and Reliance}

Mean self-reported trust (based on the HCT, which uses a 5-point Likert scale) was slightly above neutral ($M=3.22,SD=0.63)$). Participants accepted the AI's suggestions at a rate of$~.733$ (across participants: $SD=~.125$). In the 2-step setup group, when the AI suggestion disagreed with their first choice, participants changed their mind at a rate of$~.347$ (across participants: $SD=~.263$).

\begin{figure}
\centering
\begin{subfigure}{0.45\textwidth}
  \centering
  \includegraphics[width=\linewidth]{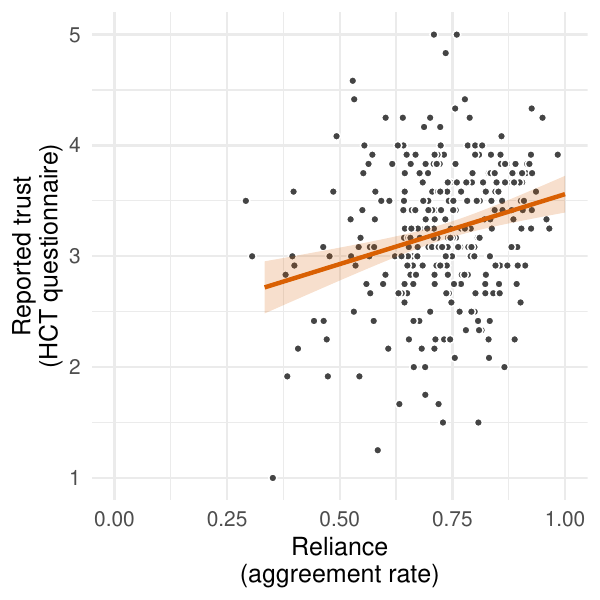}
  \caption{Agreement rate and self-reported trust are weakly to moderately correlated.}
  \label{fig:reliance_trust}
\end{subfigure}%
\hfill
\begin{subfigure}{0.45\textwidth}
  \centering
  \includegraphics[width=\linewidth]{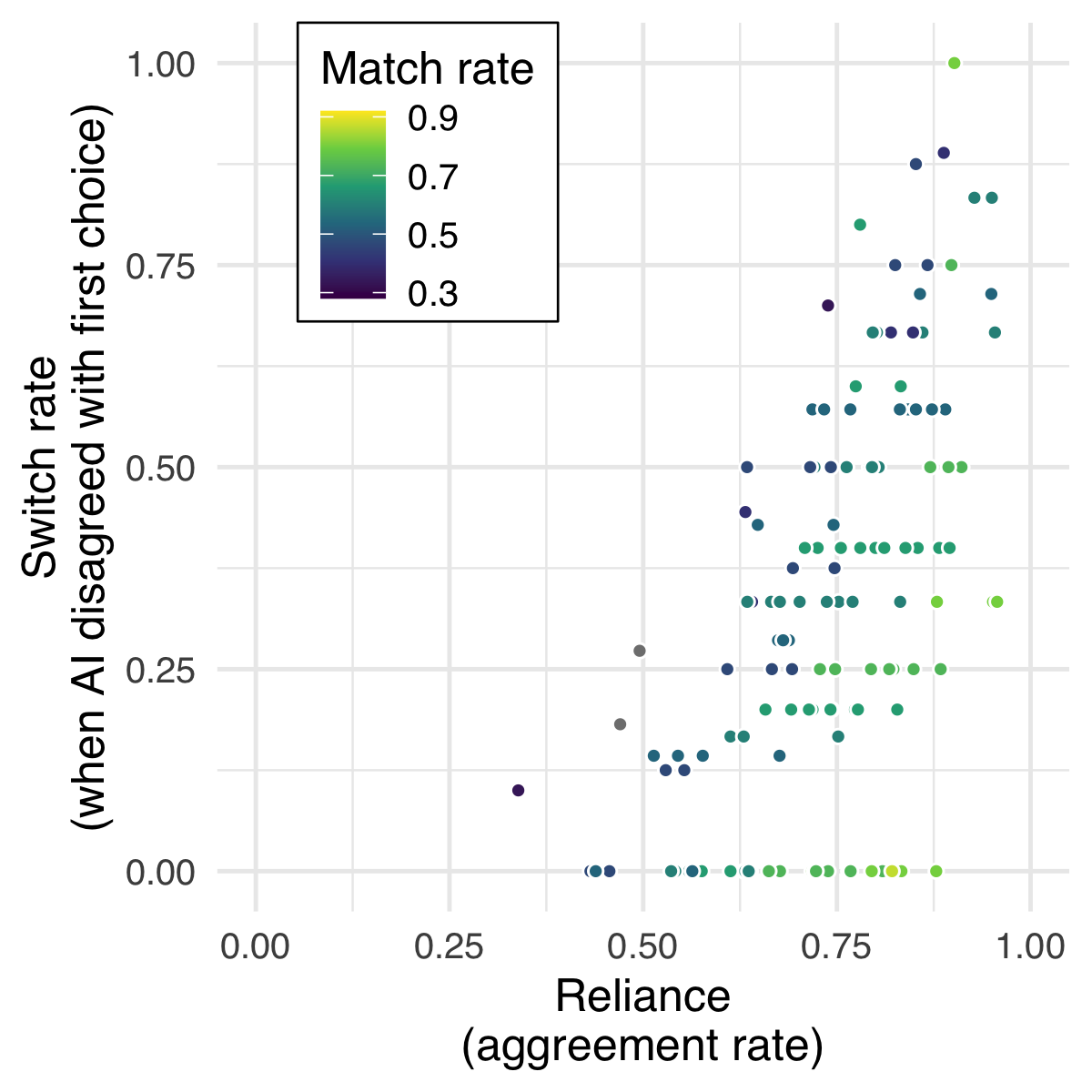}
  \caption{Agreement rate and switch rate are highly correlated. A high match means high agreement despite few switches.}
  \label{fig:reliance_switch}
\end{subfigure}
\caption{Correlation between agreement rate and reported trust/switch rate.}
\label{fig:reliance}
\end{figure}

Pearson's Correlation Coefficient was calculated to examine the relationships between the three trust/reliance measures: reported trust, overall agreement rate, and switch rate. The latter is compared for participants in the 2-step setup, and only includes cases where the AI suggestion disagreed with the participant's first choice. 
Reported trust scores did correlate positively, albeit not strongly, with the agreement rate ($r = .25$, 95\% CI $[.14, .35]$, $p < .001$) (see Figure \ref{fig:reliance_trust}) and the switch rate $(r = .25$, 95\% CI $[0.09, 0.39]$, $p < .01)$. Agreement rate and switch rate were strongly correlated ($r = .70$, 95\% CI $[0.60, 0.77]$, $p < .001$). Figure \ref{fig:reliance_switch} shows their correlation along with the \emph{match rate}, that is, the rate at which the participants' first choice matched the AI prediction. Those with a lower match rate more frequently had the opportunity to change their mind.


\subsection{Effect of Decision Setup on Reported Trust}

\begin{figure}
\centering
\begin{subfigure}{0.45\textwidth}
  \centering
  \includegraphics[width=\linewidth]{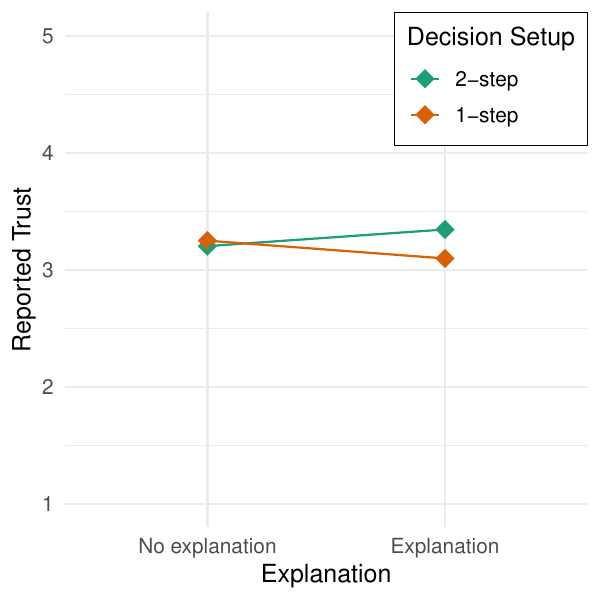}
  \caption{A 2-step setup increases reported trust, but only in the presence of an explanation.}
  \label{fig:sub1}
\end{subfigure}%
\hfill
\begin{subfigure}{0.45\textwidth}
  \centering
  \includegraphics[width=\linewidth]{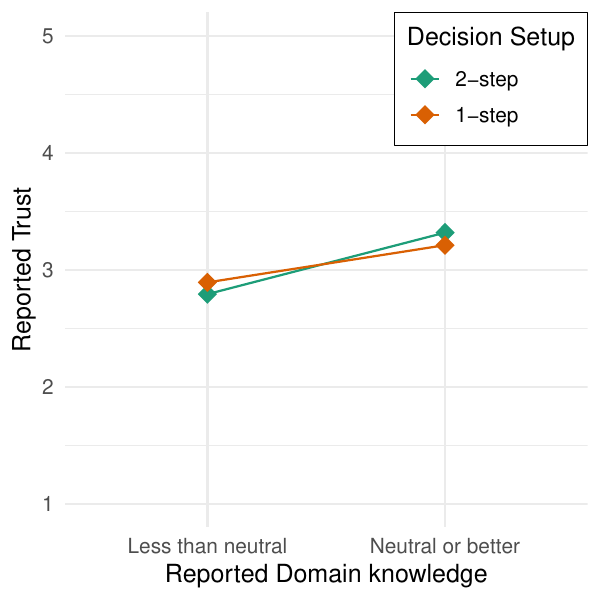}
  \caption{A 2-step setup increases reported trust only for users with at least adequate domain knowledge.}
  \label{fig:sub2}
\end{subfigure}
\caption{The decision setup (1-step or 2-step) interacts (crossover effect) both with explanations and the participants' domain knowledge, on reported trust.}
\label{fig:reported-trust}
\end{figure}

An ANCOVA linear regression was used to examine the effects of explanations and decision workflow, along with domain knowledge and prior AI experience as covariates, on reported trust (see also Figure \ref{fig:reported-trust}). The model was significant, $F(8, 289) = 3.23$, $p <~.01$, but the effect size small ($R^{2} = .08$). 

While there was no significant effect either of the explanations or of the decision workflow, there was a significant interaction between the two ($\beta = -0.50$, $p <~.01$). Participants who saw explanations reported higher trust in the 2-step group (2-step setup: $M=3.34,SD=0.61$, 1-step setup: $M=3.09,SD=0.78$), while for those who did not see explanations, it was slightly lower in the 2-step group (2-step setup: $M=3.20,SD=0.57$, 1-step setup: $M=3.25,SD=0.53$).

Domain knowledge positively predicted trust ($\beta = 0.24$, $p = .001$), and there was an interaction effect between workflow and domain knowledge ($\beta = -0.22$, $p < .05$). There was no effect of (or interaction with) prior AI experience; therefore, the final model did not include interactions with it. 

\subsection{Effect of Decision Setup on Reliance}

A generalized linear mixed model (binomial, logit link) was used to test the effects of decision setup and explanations, with covariates domain knowledge and prior AI experience, on reliance. A model was fit for general reliance (agreement rate) and 
for overreliance (agreement when the AI was wrong). 
This was on a task-by-task basis, so random intercepts for participants and tasks were included. 



For the overall agreement rate, when including all four-way interactions, there were some significant effects of explanations and interactions between domain knowledge and explanation/decision setup groups - however, these effects were not stable, and disappeared after removing the interactions with prior AI experience (which did not itself have any significant effect or interactions). Therefore, we did not include any interactions in the final model. This model did not reveal any significant main effects, however, there was a marginal effect of the decision setup ($\beta = 0.18$, $p =~.06$) on the agreement rate. 
For overreliance, no significant effects were found for explanations, domain knowledge, or AI experience, but the 2-step setup significantly increased it ($\beta = 0.46$, $p = .001$). 
Interaction effects were not significant.


\begin{figure}
\centering
\includegraphics[width=0.45\textwidth]{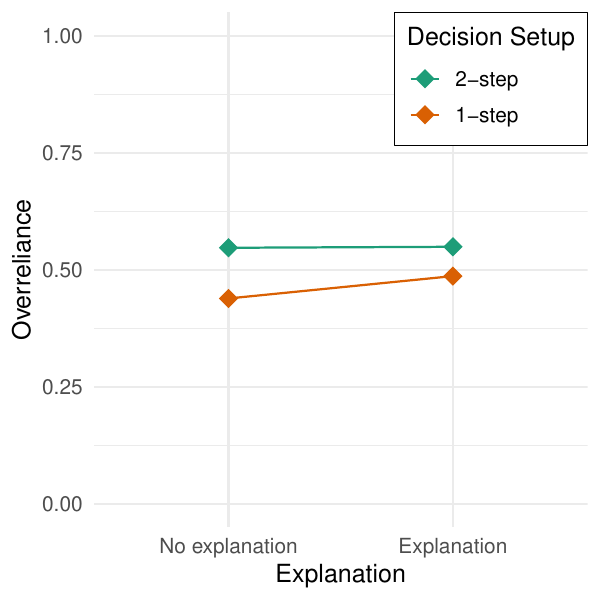}
\caption{The 2-step decision setup increases overreliance, with and without explanations.} 
\label{fig:overreliance}
\end{figure}

\section{Discussion}

\subsection{The Trust-Reliance-Paradox}


In this study, reported trust through the HCT questionnaire and reliance on the AI suggestion (agreement rate as well as switch rate) were at most moderately correlated ($r = .25$ for both reliance measures). Convergent validity between two measures is often considered adequate for correlations $> .70$ or sometimes $> .50$.  Regarding \textbf{RQ1}, we thus confirm prior suggestions that reported trust and reliance do not measure the same underlying construct \cite{rechkemmer_when_2022,lai_towards_2023}. At the same time, we did not find a clear paradox, as some have hypothesized \cite{schmitt2021towards}. As trust and reliance are to some extent related, it is more likely that they affect one another, e.g., that perceived trustworthiness of the system affects reliance together with other factors, as suggested by Schaffer et al., who showed in similar correlation in a different workflow setup \cite{schaffer_i_2019}.  


In contrast, agreement rate and switch rate were relatively strongly correlated ($r = .70$). The difference between the two measures is based on the level of complementary knowledge \cite{zhang_effect_2020,chen_understanding_2023} between the user and the AI, particularly how often the user's intuition about the outcome differs from that the AI \cite{chen_understanding_2023}, that is, the rate at which the AI prediction matches the user's first choice (\emph{match rate}): If the user and the AI are good at solving different kinds of tasks (low match rate, highly complimentary knowledge), users often have to decide between the AI prediction and their own intuition, a decision impacted by their trust in the system. If the match rate is instead higher, then the overall agreement rate can be high even with a low switch rate. Therefore, for users with a high match rate, a high agreement rate does not necessarily show trust in the AI; it might simply show that they have similar, and not complementary, outcome intuitions. Some users might report comparatively high trust because they observe that the AI predictions generally align with their own intuition - or they might report low trust in spite of high agreement rates, because they do not perceive the system to be trustworthy for other reasons. Therefore, neither reported trust nor reliance can show the whole picture on its own.

\subsection{Which Factors Impact Trust and Reliance?}


Regarding \textbf{RQ2}, we found no general (positive or negative) effect of the decision setup on reported trust independent of other factors. We did, however, find a crossover interaction effect on reported trust between the decision setup and the presence of explanations: Explanations led to higher user-reported trust in the 2-step setup, but slightly lower trust in the 1-step setup. This interaction could explain that, as prior studies and reviews have noted, results regarding the effect of explanations on trust have been contradictory \cite{schaffer_i_2019,zhang_effect_2020,bansal_does_2021,poursabzi-sangdeh_manipulating_2021,chen_understanding_2023,lai_towards_2023}. One reason for this might be that the effectiveness of explanations differs based on their type and other technical features, such as the decision workflow used.


On reliance behavior, our results do not confirm the suggestion by a prior study that a 2-step setup would decrease overreliance \cite{bucinca_2021}. Instead, here, overreliance was significantly higher in the 2-step setup, and the general agreement rate was slightly higher. Other prior studies have found that a 2-step decision workflow improves performance \cite{fogliato2022,green_2019}: this might indeed be explained by it increasing the overall agreement rate, if AI accuracy is higher than human accuracy. 


Finally, our results support arguments that users' domain knowledge needs to be accounted for in human-AI decision making tasks. Participants' domain knowledge did impact their reported trust, but not their actual reliance on the AI: Users with low domain knowledge reported significantly lower trust in the AI, although their reliance did not differ. They also reported higher trust in the 1-step setup, while those with adequate or good self-reported domain knowledge reported higher trust in the 2-step setup. Potentially, those who found it easier to assess the given data when making up their mind found the AI more trustworthy, because its prediction and explanation confirmed their own reasoning.

\subsection{Limitations and Future Work}


We simulated the AI system using pre-defined outcomes and rule-based explanations, in order to control AI accuracy to match human performance. Because of this, the explanations could not reveal systematic weaknesses of an AI system, helping users to identify sources of error. 
Moreover, the study relied on a hypothetical scenario, so that there were only imagined stakes but no real-world consequences for the participants. 
This may have led to lower engagement and impacted effects on trust and reliance. Regarding the effect of the 2-step setup in increasing overreliance, though, the results are surprising: If we had found that a 2-step decision workflow decreases overreliance, this might be explained to a certain degree by the study design - participants who did not care strongly about the outcomes of the tasks might, in the 1-step group, have simply accepted the AI suggestion without much thought, or, in the 2-step group, simply stuck to their first choice without considering the AI advice in depth. What happened, however, was that those participants in the 2-step group, who had already taken considerable time for their first choice, were afterwards more open to the AI suggestion; they were more likely to accept it, and more likely to be swayed by it even when the AI was wrong. 

As recent reviews have noted, the circumstance that many studies in AI-assisted decision making use hypothetical scenarios limits the generalizability of results \cite{lai_towards_2023,wischnewski_measuring_2023} - future studies should thus consider more realistic study designs. One next step might be to evaluate the effect of the decision setup on a realistic task with domain experts and a real AI system, as well as on tasks where the stakes are more directly felt by the participants, e.g., through performance incentives or gamification, to assess if our results can be confirmed.
Future work could further focus on more user-centered design of the decision workflow and interface, e.g., adaptive workflows, for instance by selectively employing a 2-step setup only in low-confidence cases, or giving users more control over when and what information they want to receive from the AI. 





\section{Conclusion}

We conducted an online user study ($N = 300$) on AI-assisted decision making in the domain of assigning support resources to university students. In a two-by-two between-subjects study, we compared the impact of decision workflow (1-step or 2-step setup) and explanations on users' reported and behavioral trust. 

The results of our study confirm that “trust” in AI DSS is not monolithic. Self-reported trust and behavioral reliance are only weakly related, and should be treated as distinct constructs. At the same time, our results cannot confirm that a 2-step decision workflow might improve trust calibration; instead, we found that it increased overreliance on AI advice. We also found that there was an interaction effect on reported trust between workflow and explanations: explanations raised reported trust in a 2-step decision setup, but lowered it in a 1-step setup. Moreover, self-reported trust differed based on users' domain knowledge. Therefore, we caution that results concerning the effect of explanations cannot be generalized across different study setups. 

Based on these findings, we propose implications for the design and evaluation of AI DSS. Firstly, both perceived and behavioral trust should be evaluated if possible: On the one hand, a high agreement rate can be measured when human task intuition and AI predictions match, hiding that users might perceive the system as untrustworthy. On the other hand, high trust might be reported even though users are unlikely to base their decisions on the AI advice. Thus, developers need to consider whether the goal is mainly effective collaboration, which warrants focusing on behavioral measures; or also usability or acceptability of the system, which requires accounting for user perception. Whether or not users are willing to change their minds based on AI advice can only be measured in a 2-step decision setup, however, this setup might lead to increased overreliance. Thus, a 2-step workflow might best be used to gain a better understanding of how human intuition complements the AI predictions in the given task context. It can inform system design choices, while a 1-step setup might be more suitable for later real-world applications. Developers should also consider utilizing human intuition through methods that can be adapted to the user, e.g., comparing user and system confidence, giving users the choice whether to see AI explanations, or letting them choose what kind of explanation to request. The evaluation of any chosen method should be conducted in the intended decision workflow, to ensure that experimental findings translate to later usage. 

In summary, we propose that future research focus on more adaptable explanation methods and workflows, which allow for more flexibility in helping users to integrate AI advice into their decisions. Our results suggest that one-size-fits-all solutions, such as forcing users to submit a first decision before seeing the AI prediction, are not sufficient to effectively utilize human intuition.

\begin{credits}

\subsubsection*{Acknowledgements}
This article was funded by the DFG funded Research Unit 5656 `Communicative AI',  project `P2: Interfaces' (funded by Deutsche Forschungsgemeinschaft (DFG, German Research Foundation) - 516511468) and the FET-Open Project 951846 ``MUHAI -- Meaning and Understanding for Human-centric AI'' funded by the EU program Horizon 2020.

\subsubsection{\discintname}
The authors have no competing interests to declare that are
relevant to the content of this article. 
\end{credits}
%
%
%

\bibliographystyle{splncs04}
\bibliography{refs}

@String{Computing = "Computing" }

@String{Academic = "Academic Press" }

@String{Springer = "Springer-Verlag" }

@inproceedings{jacovi_formalizing_2021,
	address = {Virtual Event Canada},
	title = {Formalizing {Trust} in {Artificial} {Intelligence}: {Prerequisites}, {Causes} and {Goals} of {Human} {Trust} in {AI}},
	isbn = {978-1-4503-8309-7},
	shorttitle = {Formalizing {Trust} in {Artificial} {Intelligence}},
	url = {https://dl.acm.org/doi/10.1145/3442188.3445923},
	doi = {10.1145/3442188.3445923},
	language = {en},
	urldate = {2023-02-24},
	booktitle = {Proceedings of the 2021 {ACM} {Conference} on {Fairness}, {Accountability}, and {Transparency}},
	publisher = {ACM},
	author = {Jacovi, Alon and Marasović, Ana and Miller, Tim and Goldberg, Yoav},
	month = mar,
	year = {2021},
	pages = {624--635},
}

@inproceedings{schmitt2021towards,
  title={Towards a trust reliance paradox? exploring the gap between perceived trust in and reliance on algorithmic advice},
  author={Schmitt, Anuschka and Wambsganss, Thiemo and S{\"o}llner, Matthias and Janson, Andreas},
  booktitle={International Conference on Information Systems (ICIS)},
  volume={1},
  pages={1--17},
  year={2021}
}

@inproceedings{he2023knowing,
  title={Knowing About Knowing: An Illusion of Human Competence Can Hinder Appropriate Reliance on AI Systems},
  author={He, Gaole and Kuiper, Lucie and Gadiraju, Ujwal},
  booktitle={Proceedings of the 2023 CHI Conference on Human Factors in Computing Systems},
  pages={1--18},
  year={2023}
}

@inproceedings{wang2021explanations,
  title={Are explanations helpful? a comparative study of the effects of explanations in ai-assisted decision-making},
  author={Wang, Xinru and Yin, Ming},
  booktitle={26th international conference on intelligent user interfaces},
  pages={318--328},
  year={2021}
}

@inproceedings{madsen2000measuring,
  title={Measuring human-computer trust},
  author={Madsen, Maria and Gregor, Shirley},
  booktitle={11th australasian conference on information systems},
  volume={53},
  pages={6--8},
  year={2000},
  organization={Citeseer}
}

@inproceedings{korber2019theoretical,
  title={Theoretical considerations and development of a questionnaire to measure trust in automation},
  author={K{\"o}rber, Moritz},
  booktitle={Proceedings of the 20th Congress of the International Ergonomics Association (IEA 2018) Volume VI: Transport Ergonomics and Human Factors (TEHF), Aerospace Human Factors and Ergonomics 20},
  pages={13--30},
  year={2019},
  organization={Springer}
}

@article{realinho2022predicting,
  title={Predicting student dropout and academic success},
  author={Realinho, Valentim and Machado, Jorge and Baptista, Lu{\'\i}s and Martins, M{\'o}nica V},
  journal={Data},
  volume={7},
  number={11},
  pages={146},
  year={2022},
  publisher={MDPI}
}

@inproceedings{bucina_proxytasks,
author = {Bu\c{c}inca, Zana and Lin, Phoebe and Gajos, Krzysztof Z. and Glassman, Elena L.},
title = {Proxy Tasks and Subjective Measures Can Be Misleading in Evaluating Explainable AI Systems},
year = {2020},
isbn = {9781450371186},
publisher = {Association for Computing Machinery},
address = {New York, NY, USA},
url = {https://doi.org/10.1145/3377325.3377498},
doi = {10.1145/3377325.3377498},
booktitle = {Proceedings of the 25th International Conference on Intelligent User Interfaces},
pages = {454–464},
numpages = {11},
location = {Cagliari, Italy},
series = {IUI '20}
}

@inproceedings{binns_reducing,
author = {Binns, Reuben and Van Kleek, Max and Veale, Michael and Lyngs, Ulrik and Zhao, Jun and Shadbolt, Nigel},
title = { 'It's Reducing a Human Being to a Percentage': Perceptions of Justice in Algorithmic Decisions},
year = {2018},
isbn = {9781450356206},
publisher = {Association for Computing Machinery},
address = {New York, NY, USA},
url = {https://doi.org/10.1145/3173574.3173951},
doi = {10.1145/3173574.3173951},
booktitle = {Proceedings of the 2018 CHI Conference on Human Factors in Computing Systems},
pages = {1–14},
numpages = {14},
location = {, Montreal QC, Canada, },
series = {CHI '18}
}

@inproceedings{spillner_eurovision,
title = {"My, my, how can I resist you?" - Examining User Reactions to Bogus Explanations of AI},
author = {Laura Spillner and Rachel Ringe and Robert Porzel and Rainer Malaka},
url = {https://ceur-ws.org/Vol-3547/paper2.pdf},
crossref = {Ethaics2023}}

@proceedings{Ethaics2023,
booktitle = {Workshop on Ethics and Trust in Human-AI Collaboration: Socio-Technical Approaches (ETHAICS 2023)},
year = 2023,
editor = {Marianna Bergamaschi Ganapini and Andrea Loreggia and Nicholas Mattei and Francesca Rossi and Biplav Srivastava and Brent Venable},
number = 3547,
series = {CEUR Workshop Proceedings},
address = {Aachen},
issn = {1613-0073},
url = {http://ceur-ws.org/Vol-3547},
venue = {Macao},
eventdate = {2023-08-21},
title = {Proceedings of the Workshop on Ethics and Trust in Human-AI Collaboration: Socio-Technical Approaches (ETHAICS 2023)}}

@article{hoffman2018metrics,
  title={Metrics for explainable AI: Challenges and prospects},
  author={Hoffman, Robert R and Mueller, Shane T and Klein, Gary and Litman, Jordan},
  journal={arXiv preprint arXiv:1812.04608},
  year={2018}
}

@inproceedings{fogliato2022,
author = {Fogliato, Riccardo and Chappidi, Shreya and Lungren, Matthew and Fisher, Paul and Wilson, Diane and Fitzke, Michael and Parkinson, Mark and Horvitz, Eric and Inkpen, Kori and Nushi, Besmira},
title = {Who Goes First? Influences of Human-AI Workflow on Decision Making in Clinical Imaging},
year = {2022},
isbn = {9781450393522},
publisher = {Association for Computing Machinery},
address = {New York, NY, USA},
url = {https://doi.org/10.1145/3531146.3533193},
doi = {10.1145/3531146.3533193},
booktitle = {Proceedings of the 2022 ACM Conference on Fairness, Accountability, and Transparency},
pages = {1362–1374},
numpages = {13},
location = {, Seoul, Republic of Korea, },
series = {FAccT '22}
}

@article{green_2019,
author = {Green, Ben and Chen, Yiling},
title = {The Principles and Limits of Algorithm-in-the-Loop Decision Making},
year = {2019},
issue_date = {November 2019},
publisher = {Association for Computing Machinery},
address = {New York, NY, USA},
volume = {3},
number = {CSCW},
url = {https://doi.org/10.1145/3359152},
doi = {10.1145/3359152},
journal = {Proc. ACM Hum.-Comput. Interact.},
month = {nov},
articleno = {50},
numpages = {24}
}

@article{bucinca_2021,
author = {Bu\c{c}inca, Zana and Malaya, Maja Barbara and Gajos, Krzysztof Z.},
title = {To Trust or to Think: Cognitive Forcing Functions Can Reduce Overreliance on AI in AI-assisted Decision-making},
year = {2021},
issue_date = {April 2021},
publisher = {Association for Computing Machinery},
address = {New York, NY, USA},
volume = {5},
number = {CSCW1},
url = {https://doi.org/10.1145/3449287},
doi = {10.1145/3449287},
journal = {Proc. ACM Hum.-Comput. Interact.},
month = {apr},
articleno = {188},
numpages = {21}
}

@misc{ma_who_2023,
	title = {Who {Should} {I} {Trust}: {AI} or {Myself}? {Leveraging} {Human} and {AI} {Correctness} {Likelihood} to {Promote} {Appropriate} {Trust} in {AI}-{Assisted} {Decision}-{Making}},
	copyright = {Creative Commons Attribution 4.0 International},
	shorttitle = {Who {Should} {I} {Trust}},
	url = {https://arxiv.org/abs/2301.05809},
	doi = {10.48550/ARXIV.2301.05809},
	urldate = {2025-08-29},
	publisher = {arXiv},
	author = {Ma, Shuai and Lei, Ying and Wang, Xinru and Zheng, Chengbo and Shi, Chuhan and Yin, Ming and Ma, Xiaojuan},
	year = {2023},
	note = {Version Number: 1},
}

@article{bucinca_trust_2021,
	title = {To {Trust} or to {Think}: {Cognitive} {Forcing} {Functions} {Can} {Reduce} {Overreliance} on {AI} in {AI}-assisted {Decision}-making},
	volume = {5},
	issn = {2573-0142},
	shorttitle = {To {Trust} or to {Think}},
	url = {https://dl.acm.org/doi/10.1145/3449287},
	doi = {10.1145/3449287},
	language = {en},
	number = {CSCW1},
	urldate = {2025-08-29},
	journal = {Proceedings of the ACM on Human-Computer Interaction},
	author = {Buçinca, Zana and Malaya, Maja Barbara and Gajos, Krzysztof Z.},
	month = apr,
	year = {2021},
	pages = {1--21},
}

@misc{chen_understanding_2023,
	title = {Understanding the {Role} of {Human} {Intuition} on {Reliance} in {Human}-{AI} {Decision}-{Making} with {Explanations}},
	copyright = {Creative Commons Attribution 4.0 International},
	url = {https://arxiv.org/abs/2301.07255},
	doi = {10.48550/ARXIV.2301.07255},
	urldate = {2025-08-29},
	publisher = {arXiv},
	author = {Chen, Valerie and Liao, Q. Vera and Vaughan, Jennifer Wortman and Bansal, Gagan},
	year = {2023},
	note = {Version Number: 3},
}

@inproceedings{lai_towards_2023,
	address = {Chicago IL USA},
	title = {Towards a {Science} of {Human}-{AI} {Decision} {Making}: {An} {Overview} of {Design} {Space} in {Empirical} {Human}-{Subject} {Studies}},
	isbn = {9798400701924},
	shorttitle = {Towards a {Science} of {Human}-{AI} {Decision} {Making}},
	url = {https://dl.acm.org/doi/10.1145/3593013.3594087},
	doi = {10.1145/3593013.3594087},
	language = {en},
	urldate = {2025-08-29},
	booktitle = {2023 {ACM} {Conference} on {Fairness} {Accountability} and {Transparency}},
	publisher = {ACM},
	author = {Lai, Vivian and Chen, Chacha and Smith-Renner, Alison and Liao, Q. Vera and Tan, Chenhao},
	month = jun,
	year = {2023},
	pages = {1369--1385},
}

@inproceedings{rechkemmer_when_2022,
	address = {New Orleans LA USA},
	title = {When {Confidence} {Meets} {Accuracy}: {Exploring} the {Effects} of {Multiple} {Performance} {Indicators} on {Trust} in {Machine} {Learning} {Models}},
	copyright = {https://www.acm.org/publications/policies/copyright\_policy\#Background},
	isbn = {978-1-4503-9157-3},
	shorttitle = {When {Confidence} {Meets} {Accuracy}},
	url = {https://dl.acm.org/doi/10.1145/3491102.3501967},
	doi = {10.1145/3491102.3501967},
	language = {en},
	urldate = {2025-08-29},
	booktitle = {{CHI} {Conference} on {Human} {Factors} in {Computing} {Systems}},
	publisher = {ACM},
	author = {Rechkemmer, Amy and Yin, Ming},
	month = apr,
	year = {2022},
	pages = {1--14},
}

@article{banovic_being_2023,
	title = {Being {Trustworthy} is {Not} {Enough}: {How} {Untrustworthy} {Artificial} {Intelligence} ({AI}) {Can} {Deceive} the {End}-{Users} and {Gain} {Their} {Trust}},
	volume = {7},
	issn = {2573-0142},
	shorttitle = {Being {Trustworthy} is {Not} {Enough}},
	url = {https://dl.acm.org/doi/10.1145/3579460},
	doi = {10.1145/3579460},
	language = {en},
	number = {CSCW1},
	urldate = {2025-08-31},
	journal = {Proceedings of the ACM on Human-Computer Interaction},
	author = {Banovic, Nikola and Yang, Zhuoran and Ramesh, Aditya and Liu, Alice},
	month = apr,
	year = {2023},
	pages = {1--17},
}

@inproceedings{wazzan_evaluating_2025,
	address = {Cagliari Italy},
	title = {Evaluating the {Impact} of {AI}-{Generated} {Visual} {Explanations} on {Decision}-{Making} for {Image} {Matching}},
	isbn = {9798400713064},
	url = {https://dl.acm.org/doi/10.1145/3708359.3712121},
	doi = {10.1145/3708359.3712121},
	language = {en},
	urldate = {2025-08-31},
	booktitle = {Proceedings of the 30th {International} {Conference} on {Intelligent} {User} {Interfaces}},
	publisher = {ACM},
	author = {Wazzan, Albatool and Wright, Marcus and MacNeil, Stephen and Souvenir, Richard},
	month = mar,
	year = {2025},
	pages = {672--684},
}

@inproceedings{he_is_2025,
	address = {Cagliari Italy},
	title = {Is {Conversational} {XAI} {All} {You} {Need}? {Human}-{AI} {Decision} {Making} {With} a {Conversational} {XAI} {Assistant}},
	isbn = {9798400713064},
	shorttitle = {Is {Conversational} {XAI} {All} {You} {Need}?},
	url = {https://dl.acm.org/doi/10.1145/3708359.3712133},
	doi = {10.1145/3708359.3712133},
	language = {en},
	urldate = {2025-08-31},
	booktitle = {Proceedings of the 30th {International} {Conference} on {Intelligent} {User} {Interfaces}},
	publisher = {ACM},
	author = {He, Gaole and Aishwarya, Nilay and Gadiraju, Ujwal},
	month = mar,
	year = {2025},
	pages = {907--924},
}

@inproceedings{li_as_2025,
	address = {Yokohama Japan},
	title = {As {Confidence} {Aligns}: {Understanding} the {Effect} of {AI} {Confidence} on {Human} {Self}-confidence in {Human}-{AI} {Decision} {Making}},
	isbn = {9798400713941},
	shorttitle = {As {Confidence} {Aligns}},
	url = {https://dl.acm.org/doi/10.1145/3706598.3713336},
	doi = {10.1145/3706598.3713336},
	language = {en},
	urldate = {2025-08-31},
	booktitle = {Proceedings of the 2025 {CHI} {Conference} on {Human} {Factors} in {Computing} {Systems}},
	publisher = {ACM},
	author = {Li, Jingshu and Yang, Yitian and Liao, Q. Vera and Zhang, Junti and Lee, Yi-Chieh},
	month = apr,
	year = {2025},
	pages = {1--16},
}

@misc{xiong_can_2023,
	title = {Can {LLMs} {Express} {Their} {Uncertainty}? {An} {Empirical} {Evaluation} of {Confidence} {Elicitation} in {LLMs}},
	copyright = {arXiv.org perpetual, non-exclusive license},
	shorttitle = {Can {LLMs} {Express} {Their} {Uncertainty}?},
	url = {https://arxiv.org/abs/2306.13063},
	doi = {10.48550/ARXIV.2306.13063},
	urldate = {2025-08-31},
	publisher = {arXiv},
	author = {Xiong, Miao and Hu, Zhiyuan and Lu, Xinyang and Li, Yifei and Fu, Jie and He, Junxian and Hooi, Bryan},
	year = {2023},
	note = {Version Number: 2},
}

@inproceedings{zhang_effect_2020,
	address = {Barcelona Spain},
	title = {Effect of confidence and explanation on accuracy and trust calibration in {AI}-assisted decision making},
	isbn = {978-1-4503-6936-7},
	url = {https://dl.acm.org/doi/10.1145/3351095.3372852},
	doi = {10.1145/3351095.3372852},
	language = {en},
	urldate = {2025-08-31},
	booktitle = {Proceedings of the 2020 {Conference} on {Fairness}, {Accountability}, and {Transparency}},
	publisher = {ACM},
	author = {Zhang, Yunfeng and Liao, Q. Vera and Bellamy, Rachel K. E.},
	month = jan,
	year = {2020},
	pages = {295--305},
}

@inproceedings{kunkel_let_2019,
	address = {Glasgow Scotland Uk},
	title = {Let {Me} {Explain}: {Impact} of {Personal} and {Impersonal} {Explanations} on {Trust} in {Recommender} {Systems}},
	isbn = {978-1-4503-5970-2},
	shorttitle = {Let {Me} {Explain}},
	url = {https://dl.acm.org/doi/10.1145/3290605.3300717},
	doi = {10.1145/3290605.3300717},
	language = {en},
	urldate = {2025-08-31},
	booktitle = {Proceedings of the 2019 {CHI} {Conference} on {Human} {Factors} in {Computing} {Systems}},
	publisher = {ACM},
	author = {Kunkel, Johannes and Donkers, Tim and Michael, Lisa and Barbu, Catalin-Mihai and Ziegler, Jürgen},
	month = may,
	year = {2019},
	pages = {1--12},
}

@inproceedings{suresh_misplaced_2020,
	address = {Southampton United Kingdom},
	title = {Misplaced {Trust}: {Measuring} the {Interference} of {Machine} {Learning} in {Human} {Decision}-{Making}},
	isbn = {978-1-4503-7989-2},
	shorttitle = {Misplaced {Trust}},
	url = {https://dl.acm.org/doi/10.1145/3394231.3397922},
	doi = {10.1145/3394231.3397922},
	language = {en},
	urldate = {2025-08-31},
	booktitle = {12th {ACM} {Conference} on {Web} {Science}},
	publisher = {ACM},
	author = {Suresh, Harini and Lao, Natalie and Liccardi, Ilaria},
	month = jul,
	year = {2020},
	pages = {315--324},
}

@inproceedings{wang_are_2021,
	address = {College Station TX USA},
	title = {Are {Explanations} {Helpful}? {A} {Comparative} {Study} of the {Effects} of {Explanations} in {AI}-{Assisted} {Decision}-{Making}},
	isbn = {978-1-4503-8017-1},
	shorttitle = {Are {Explanations} {Helpful}?},
	url = {https://dl.acm.org/doi/10.1145/3397481.3450650},
	doi = {10.1145/3397481.3450650},
	language = {en},
	urldate = {2025-08-31},
	booktitle = {26th {International} {Conference} on {Intelligent} {User} {Interfaces}},
	publisher = {ACM},
	author = {Wang, Xinru and Yin, Ming},
	month = apr,
	year = {2021},
	pages = {318--328},
}

@inproceedings{poursabzi-sangdeh_manipulating_2021,
	address = {Yokohama Japan},
	title = {Manipulating and {Measuring} {Model} {Interpretability}},
	isbn = {978-1-4503-8096-6},
	url = {https://dl.acm.org/doi/10.1145/3411764.3445315},
	doi = {10.1145/3411764.3445315},
	language = {en},
	urldate = {2025-08-31},
	booktitle = {Proceedings of the 2021 {CHI} {Conference} on {Human} {Factors} in {Computing} {Systems}},
	publisher = {ACM},
	author = {Poursabzi-Sangdeh, Forough and Goldstein, Daniel G and Hofman, Jake M and Wortman Vaughan, Jennifer Wortman and Wallach, Hanna},
	month = may,
	year = {2021},
	pages = {1--52},
}

@inproceedings{schaffer_i_2019,
	address = {Marina del Ray California},
	title = {I can do better than your {AI}: expertise and explanations},
	isbn = {978-1-4503-6272-6},
	shorttitle = {I can do better than your {AI}},
	url = {https://dl.acm.org/doi/10.1145/3301275.3302308},
	doi = {10.1145/3301275.3302308},
	language = {en},
	urldate = {2025-08-31},
	booktitle = {Proceedings of the 24th {International} {Conference} on {Intelligent} {User} {Interfaces}},
	publisher = {ACM},
	author = {Schaffer, James and O'Donovan, John and Michaelis, James and Raglin, Adrienne and Höllerer, Tobias},
	month = mar,
	year = {2019},
	pages = {240--251},
}

@article{bach_systematic_2024,
	title = {A {Systematic} {Literature} {Review} of {User} {Trust} in {AI}-{Enabled} {Systems}: {An} {HCI} {Perspective}},
	volume = {40},
	issn = {1044-7318, 1532-7590},
	shorttitle = {A {Systematic} {Literature} {Review} of {User} {Trust} in {AI}-{Enabled} {Systems}},
	url = {https://www.tandfonline.com/doi/full/10.1080/10447318.2022.2138826},
	doi = {10.1080/10447318.2022.2138826},
	language = {en},
	number = {5},
	urldate = {2025-08-31},
	journal = {International Journal of Human–Computer Interaction},
	author = {Bach, Tita Alissa and Khan, Amna and Hallock, Harry and Beltrão, Gabriela and Sousa, Sonia},
	month = mar,
	year = {2024},
	pages = {1251--1266},
}

@inproceedings{bansal_does_2021,
	address = {Yokohama Japan},
	title = {Does the {Whole} {Exceed} its {Parts}? {The} {Effect} of {AI} {Explanations} on {Complementary} {Team} {Performance}},
	isbn = {978-1-4503-8096-6},
	shorttitle = {Does the {Whole} {Exceed} its {Parts}?},
	url = {https://dl.acm.org/doi/10.1145/3411764.3445717},
	doi = {10.1145/3411764.3445717},
	language = {en},
	urldate = {2025-09-02},
	booktitle = {Proceedings of the 2021 {CHI} {Conference} on {Human} {Factors} in {Computing} {Systems}},
	publisher = {ACM},
	author = {Bansal, Gagan and Wu, Tongshuang and Zhou, Joyce and Fok, Raymond and Nushi, Besmira and Kamar, Ece and Ribeiro, Marco Tulio and Weld, Daniel},
	month = may,
	year = {2021},
	pages = {1--16},
}

@article{wang_effects_2022,
	title = {Effects of {Explanations} in {AI}-{Assisted} {Decision} {Making}: {Principles} and {Comparisons}},
	volume = {12},
	issn = {2160-6455, 2160-6463},
	shorttitle = {Effects of {Explanations} in {AI}-{Assisted} {Decision} {Making}},
	doi = {10.1145/3519266},
	language = {en},
	journal = {ACM Transactions on Interactive Intelligent Systems},
	author = {Wang, Xinru and Yin, Ming},
	year = {2022},
}

@article{huang_survey_2025,
	title = {A {Survey} on {Hallucination} in {Large} {Language} {Models}: {Principles}, {Taxonomy}, {Challenges}, and {Open} {Questions}},
	volume = {43},
	issn = {1046-8188, 1558-2868},
	shorttitle = {A {Survey} on {Hallucination} in {Large} {Language} {Models}},
	doi = {10.1145/3703155},
	journal = {ACM Transactions on Information Systems},
	author = {Huang, Lei and Yu, Weijiang and Ma, Weitao and Zhong, Weihong and Feng, Zhangyin and Wang, Haotian and Chen, Qianglong and Peng, Weihua and Feng, Xiaocheng and Qin, Bing and Liu, Ting},
	year = {2025},
}

@inproceedings{wischnewski_measuring_2023,
	address = {Hamburg Germany},
	title = {Measuring and {Understanding} {Trust} {Calibrations} for {Automated} {Systems}: {A} {Survey} of the {State}-{Of}-{The}-{Art} and {Future} {Directions}},
	isbn = {978-1-4503-9421-5},
	shorttitle = {Measuring and {Understanding} {Trust} {Calibrations} for {Automated} {Systems}},
	doi = {10.1145/3544548.3581197},
	booktitle = {Proceedings of the 2023 {CHI} {Conference} on {Human} {Factors} in {Computing} {Systems}},
	publisher = {ACM},
	author = {Wischnewski, Magdalena and Krämer, Nicole and Müller, Emmanuel},
	year = {2023},
}

\end{document}